\documentclass[article,preprint2,onecolappendix]{aastex61}

  \usepackage{amsmath}
  \usepackage{tabularx}
  \usepackage{color}
  \usepackage{graphicx}
  \usepackage{rotating}
  \usepackage{subfigure}

\hypersetup{linkcolor=red,citecolor=blue,filecolor=cyan,urlcolor=magenta}

\newcommand\aastex{AAS\TeX}

\received{February, 2018}
\revised{\today}
\accepted{2018}
\submitjournal{ApJL}

\shorttitle{\aastex\ }
\shortauthors{Paiano et al.}

\begin{document}

\title{The redshift of the BL Lac object TXS~0506+056.\\}

\correspondingauthor{Simona Paiano}
\email{simona.paiano@oapd.inaf.it}

\author{Simona Paiano}
\affiliation{INAF, Osservatorio Astronomico di Padova, Vicolo dell'Osservatorio 5 I-35122 Padova - ITALY}
\affiliation{INFN, Sezione di Padova, via Marzolo 8, I-35131 Padova - ITALY}

\author{Renato Falomo}
\affiliation{INAF, Osservatorio Astronomico di Padova, Vicolo dell'Osservatorio 5 I-35122 Padova - ITALY}

\author{Aldo Treves}
\affiliation{Universit\`a degli Studi dell'Insubria, Via Valleggio 11 I-22100 Como - ITALY}
\affiliation{INAF, Osservatorio Astronomico di Brera, Via E. Bianchi 46 I-23807 Merate (LC) - ITALY}

\author{Riccardo Scarpa}
\affiliation{Instituto de Astrofisica de Canarias, C/O Via Lactea, s/n E38205 - La Laguna (Tenerife) - SPAIN}
\affiliation{Universidad de La Laguna, Dpto. Astrofisica, s/n E-38206 La Laguna (Tenerife) - SPAIN}

\begin{abstract}
The bright BL~Lac object TXS~0506+056 is a most likely counterpart of the IceCube neutrino event EHE~170922A. 
The lack of this redshift prevents  a comprehensive understanding of the modeling of the source. 
We present high signal-to-noise optical spectroscopy, in the range 4100-9000~$\textrm{\AA}$, obtained at the 10.4m Gran Telescopio Canarias. 
The spectrum is characterized by a power law continuum and is marked by faint interstellar features. 
In the regions unaffected by these features, we found three very weak (EW~$\sim$~0.1~$\textrm{\AA}$) emission lines that we identify with [O~II]~3727~$\textrm{\AA}$,  [O~III]~~5007~$\textrm{\AA}$, and [NII]~6583~$\textrm{\AA}$, yielding the redshift z~=~0.3365$\pm$0.0010.
\end{abstract}


\section{Introduction} \label{sec:intro}

The radio source TXS~0506+056 (3FGL J0509.4+0541) is a bright BL Lac object (V~$\sim$~15), that recently became of the utmost astrophysical interest since it is considered as the probable counterpart of the IceCube neutrino event EHE~170922A of 22 September 2017 \citep{atel-icecube2017}. 
This event is believed to be associated to enhanced $\gamma$-ray  (100~MeV~-~300~GeV) activity of the \textit{Fermi}/LAT source 3FGL~J0509.4+0541 \citep{atel-fermi2017, atel-agile2017} and to a significant detection at $>$~100~GeV by the MAGIC telescopes \citep{atel-magic2017}. 
An enhanced flux was also revealed in the X-ray regime by the \textit{Swift} satellite \citep{atel-swift2017} and from the ASAS-SN survey in the optical band \citep{atel-asassn2017}. 
During the event, the optical monitoring of the source indicates that the object was $\sim$0.5 mag brighter in the V band with respect to the earlier months. 

Optical spectroscopy obtained in previous years by \citet{halpern2003}, \citet{shaw2013a} and \citet{landoni2013}, and after  the neutrino event, failed to determine the redshift of this object \citep{atel-liverpool2017, atel-salt2017,  atel-vlt2017, atel-subaru2017}.

Since the knowledge of the TXS~0506+056 distance is mandatory to model the spectral energy distribution and the neutrino production mechanism, we undertook a detailed spectroscopic study at the 10.4m Gran Telescopio Canarias (GTC).

Here we report the results obtained from high signal-to-noise optical observations aimed to pin down the redshift of the source.

\section{Observations and data analysis} \label{sec:floats}

\begin{table}
\caption{LOG OF THE OBSERVATIONS}\label{tab:table2}
\centering
\begin{tabular}{lccc}
\hline 
Grism    &  Date  &  Total exp.    &  N   \\
             &            &   time (s)      &         \\
\hline
R1000B   & 23-11-2017   & 3600  &  5\\
                &  05-12-2017  & 4200 &   6 \\
\hline
R1000R  & 02-01-2018    & 4000 &  6 \\
               & 14-01-2018    & 4000 &  6 \\
\hline
R2500V  & 14-01-2018  & 4800  &   3\\
               & 14-01-2018  & 4800  &   3\\
\hline
R2500R  & 15-01-2018    & 4500 & 3\\
               & 20-01-2018    & 4800 &  6 \\
\hline
R2500I   & 10-01-2018   & 4500  & 3 \\
              & 13-01-2018   & 4500  &  2  \\
              & 20-01-2018   & 4800  &  6 \\
\hline
\end{tabular}
\tablenotetext{}{
\raggedright
\footnotesize \texttt{Col.1}: Grism name (slit width~=~1.0" for R1000 and slit width~=~1.2" for R2500); \texttt{Col.2}: Date of the observation, \texttt{Col.3}: Total exposure time, \texttt{Col.4}: Number of individual exposures.}
\tablenotetext{}{
\raggedright} 
\end{table}

TXS~0506+056 was observed with the GTC at the Roque de Los Muchachos with the spectrograph OSIRIS \citep{cepa2003} covering the spectral range 4100~-~9000~$\textrm{\AA}$. 
We adopted different grisms (R1000 and R2500) yielding spectral resolution R~=~$\lambda / \Delta\lambda \sim$~600 and $\sim$~1300.
For each setting, we obtained many independent exposures (see Tab. 1).

Data reduction was carried out using IRAF software and standard procedures for long slit spectroscopy, following the same scheme given in \citet{paiano2017tev}. 
The accuracy of the wavelength calibration is 0.1~$\textrm{\AA}$. 
Relative flux calibration was derived from the observations of spectro-photometric standard stars. 

For each dataset, we combined the independent exposures, weighting for signal-to-noise ratio (S/N), and anchored the absolute flux to the averaged magnitude of the source (g~=~15.4), as from the acquisition image.
Finally the spectra were dereddened applying the extinction law by \citet{cardelli1989}, assuming  E$_{B-V}$~=~0.1
as from the NASA/IPAC Infrared Science Archive 6\footnote{http://irsa.ipac.caltech.edu/applications/DUST/} . 

\begin{figure*}
\includegraphics[width=0.8\textwidth, angle=-90]{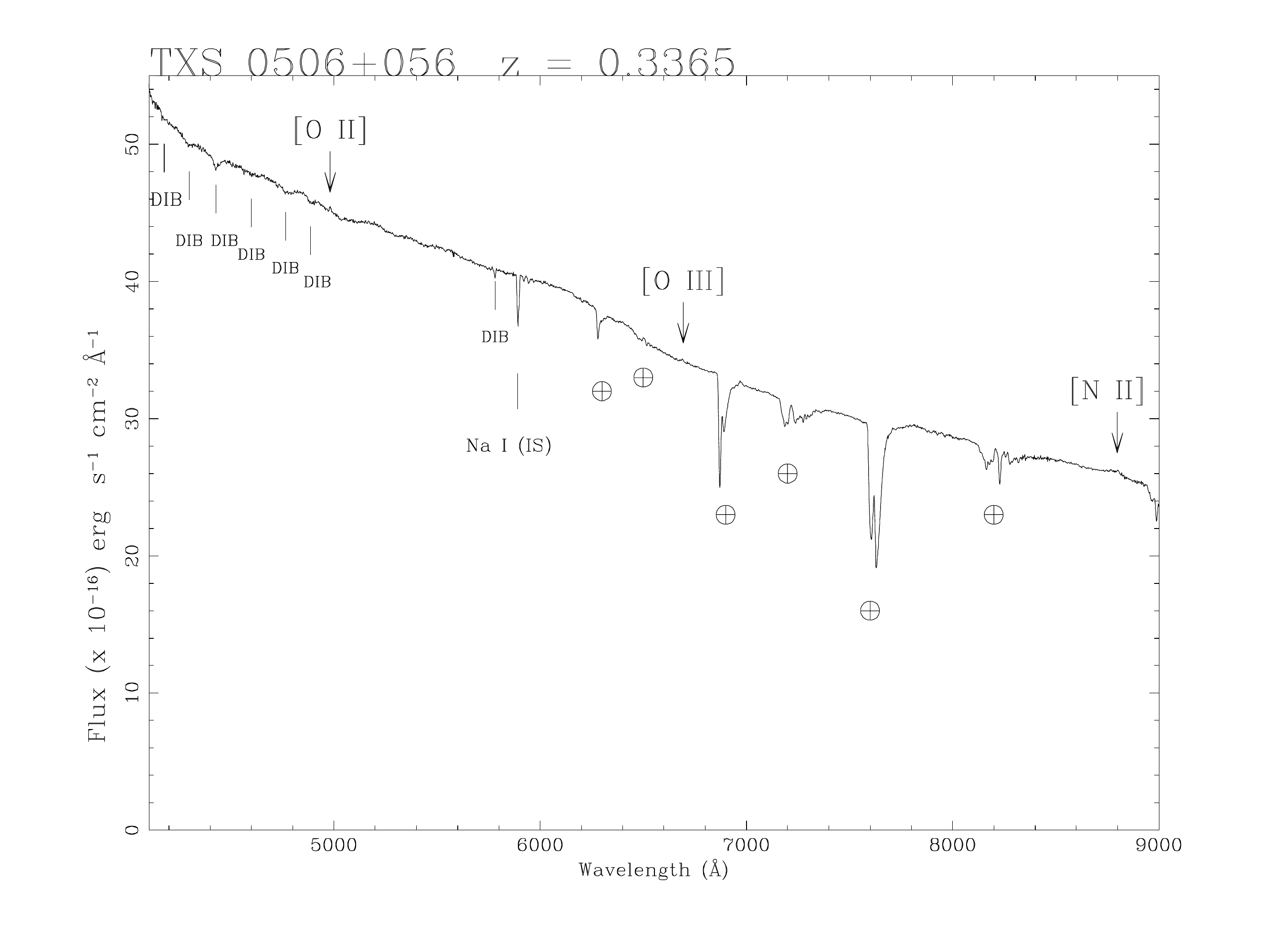}
\caption{The optical spectrum of the BL Lac object TXS~0506+056 obtained at GTC+OSIRIS (R$\sim$600).  The spectrum is corrected for reddening assuming E$_{B-V}$~=~0.1. The shape of the spectrum is dominated by a non-thermal emission and the three arrows indicate the position of the detected emission lines at z~=~0.3365. Absorption features due to interstellar medium are labelled as DIB and IS. The main telluric bands are marked by $\oplus$.}
\label{fig:fig1}
\end{figure*}

\section{Results} \label{subsec:tables}

In Fig.~\ref{fig:fig1} we show the full R$\sim$600 optical spectrum. The S/N ranges from 650 to 1200 depending on the wavelength.
The spectrum is characterized by a non-thermal emission with the power law shape (F$_{\lambda} \propto \lambda^{\alpha}$) with spectral index of $\alpha$~=~-1.0~$\pm$~0.1.

Beside the prominent telluric absorptions and interstellar features, the whole optical spectrum does not exhibit emission or absorption lines with equivalent width~$>$~0.5~$\textrm{\AA}$.

\begin{table}
\caption{Optical spectral features }\label{tab:table2}
\centering
\begin{tabular}{lll}
\hline 
$\lambda$  ($\textrm{\AA}$)   &  EW   ($\textrm{\AA}$)   & ID\\
\hline
4190   &  0.20 $\pm$ 0.05    & DIB \\
4290   &  0.15 $\pm$ 0.05    & DIB \\
4427   &  1.00 $\pm$ 0.10    & DIB \\
4600   &  0.30 $\pm$ 0.07    & DIB \\
4770   &  0.35 $\pm$ 0.05    & DIB \\
4890   &  0.30 $\pm$ 0.06    & DIB \\
5780   &  0.35 $\pm$ 0.03    & DIB \\
\hline
\hline
4981.5  &   0.12 $\pm$ 0.03 &    [OII] 3727$\textrm{\AA}$\\
6693.6  &   0.17 $\pm$ 0.02 &   [OIII] 5007$\textrm{\AA}$\\
8800.5  &   0.05 $\pm$ 0.02 &  [NII]  6583$\textrm{\AA}$\\
\hline
\end{tabular}
\tablenotetext{}{
\raggedright
\footnotesize \texttt{Col.1}: Central wavelength of the feature; \texttt{Col.2}: Observed equivalent width of the feature, \texttt{Col.3}: Feature identification.}
\tablenotetext{}{
\raggedright
 } 
\end{table}

\begin{figure*}
\includegraphics[width=0.7\textwidth, angle=-90]{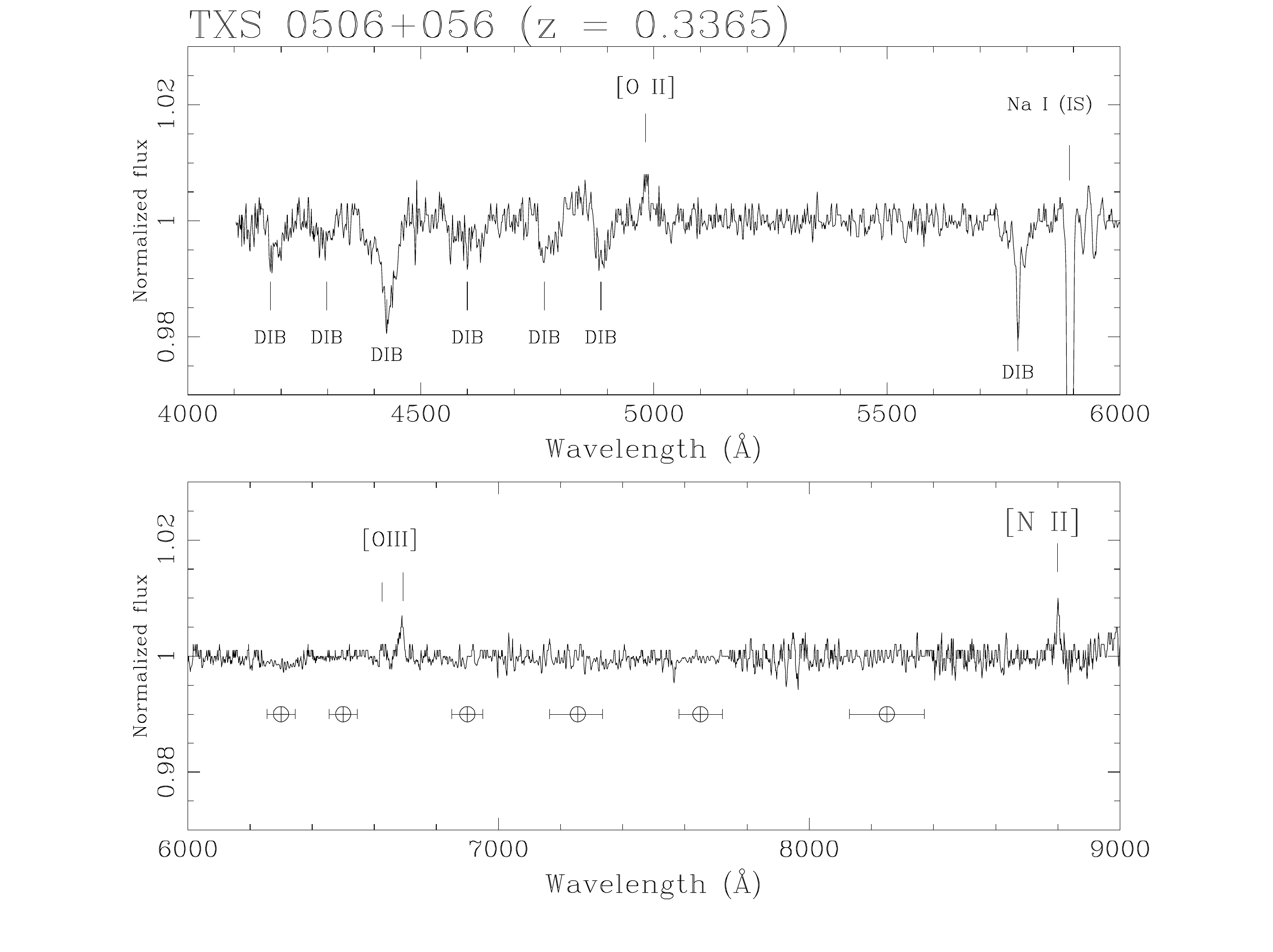}
\caption{The normalized spectrum of the BL Lac object TXS~0506+056 (see also Fig.1). Three weak (EW$\sim$~0.1~$\textrm{\AA}$) emission lines are detected and identified as [OII]~3727~$\textrm{\AA}$, [OIII] 5007$\textrm{\AA}$, and [NII]  6583$\textrm{\AA}$ at the redshift z~=~0.3365. Absorption features due to interstellar medium are labelled as DIB and IS. The spectral region affected by telluric absorptions, indicated by $\oplus$, were corrected. }
\label{fig:fig2}
\end{figure*}

In order to evidentiate the spectral features, we divided it by a fit of the continuum, obtained excluding all regions affected by atmospheric and interstellar absorptions (see Fig.~\ref{fig:fig2}).
From the normalized spectrum, we computed the nominal equivalent width (EW) in five intervals of the spectra, avoiding the prominent telluric absorption features (see details about the used procedure in Appendix A in \citet{paiano2017tev}). 
Five different intervals were considered because the S/N depends on the wavelength. 
This translates into a minimum (3$\sigma$~level) detectable equivalent width: EW$_{min}$~=~0.05~-~0.1.

The spectral region between 4100~$\textrm{\AA}$ and 4900~$\textrm{\AA}$ is affected by Diffuse Interstellar Bands (DIBs) \citep[e.g.][]{herbig1995} of EW~=~0.3~-~1.0~$\textrm{\AA}$ (see Table 2).
In addition, the prominent (EW~=~1.2~$\textrm{\AA}$) interstellar absorption line due to Na~I~5892 $\textrm{\AA}$ is found\footnote{The source is at Galactic latitude $l~=~-19.6^{\circ}$}.

We search for weak intrinsic absorption and/or emission lines in the spectrum, avoiding all the interstellar spectral features and the regions clearly dominated by the telluric absorptions due to O$_2$ and H$_2$O.
We detect three faint narrow emission lines at 4981.5~$\pm$~1.0~$\textrm{\AA}$, 6693.6~$\pm$~1.1~$\textrm{\AA}$ and  at 8800.5~$\pm$~1.1~$\textrm{\AA}$ (see Fig.~\ref{fig:fig2}), identified as [OII]~3727~$\textrm{\AA}$, [OIII]~5007~$\textrm{\AA}$, and [NII]~6583~$\textrm{\AA}$, respectively, at z~=~0.336.

\begin{figure}
\centering
\includegraphics[bb = 30 180 400 510, width=0.37\textwidth]{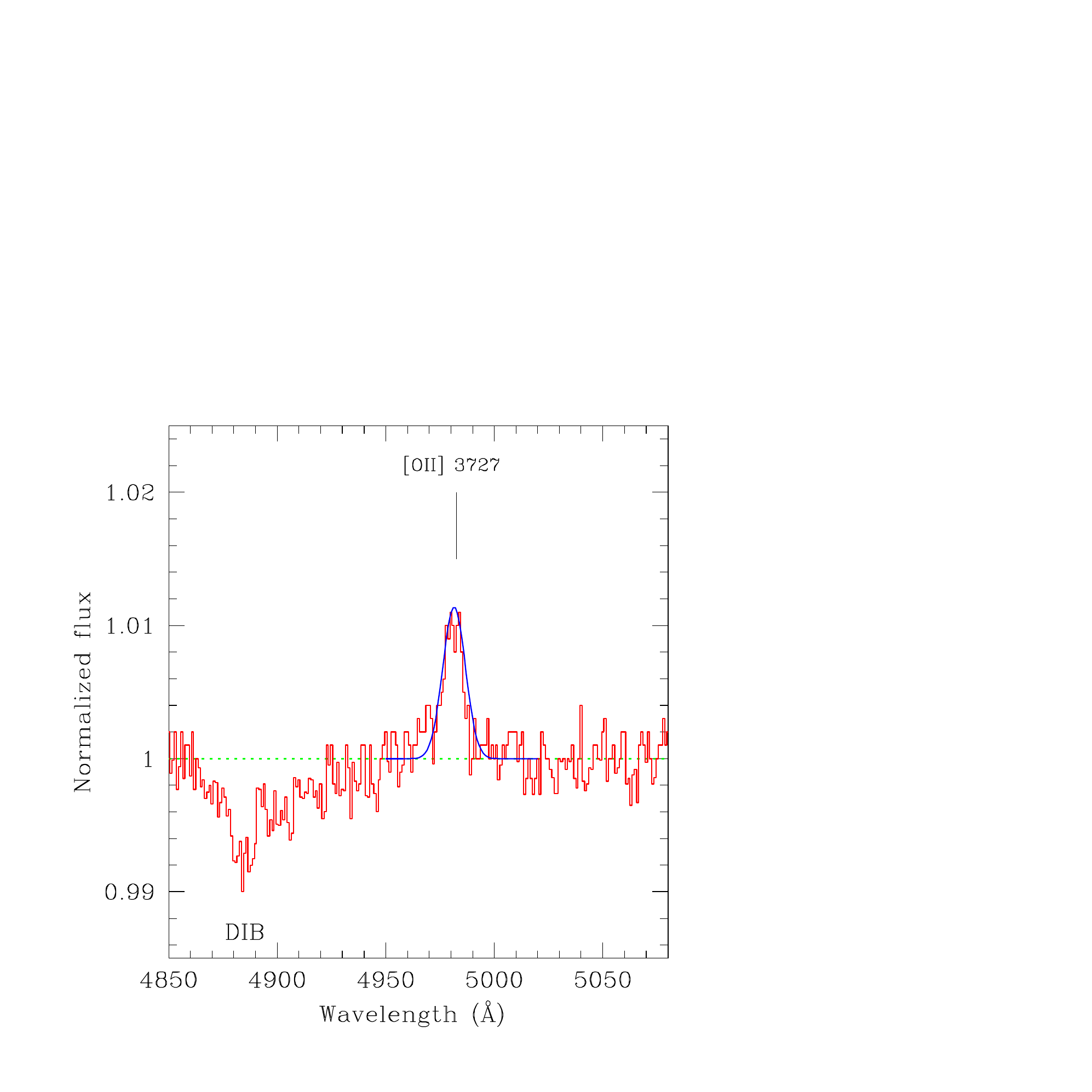}
\includegraphics[bb = 30 180 400 510, width=0.37\textwidth]{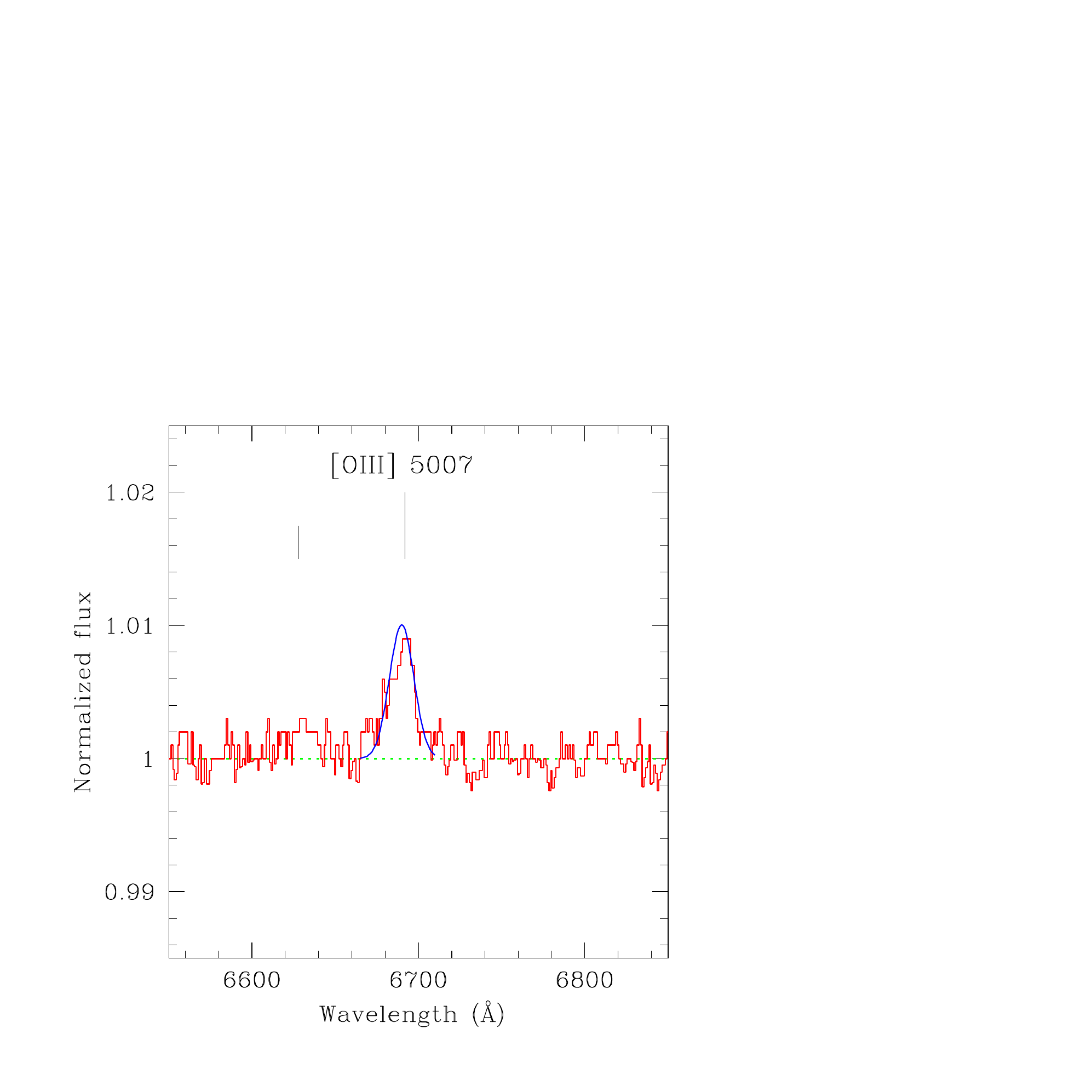}
\includegraphics[bb = 30 180 400 510, width=0.37\textwidth]{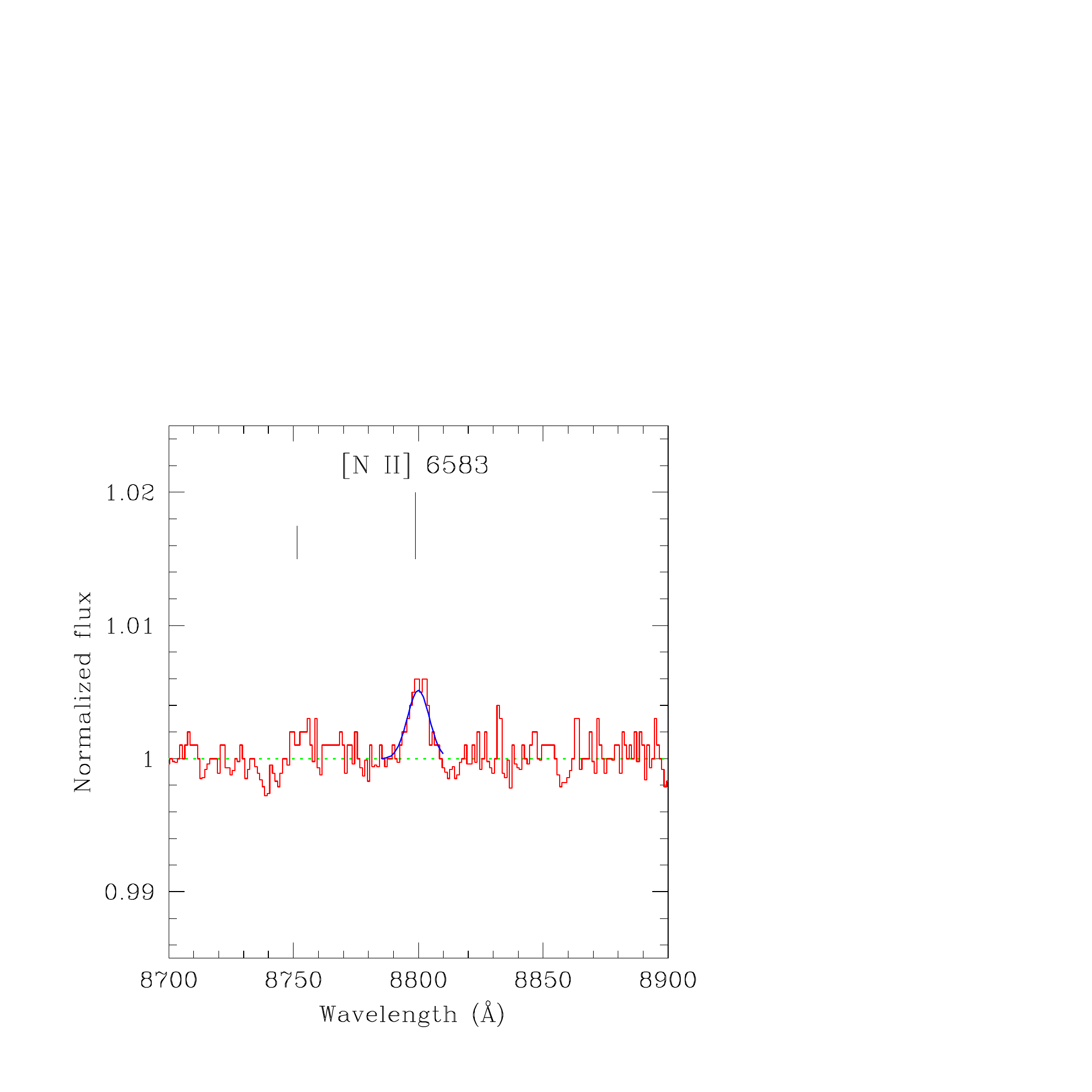}
\vspace{2.5cm}
\caption{Close up of the normalized optical spectrum (obtained with R$\sim$1300 ) of TXS~0506+056 around the three faint detected emission lines. \textit{Top}: The emission line at 4981.5~$\textrm{\AA}$ identified as [OII]~3727~$\textrm{\AA}$ (EW~=~0.12~$\textrm{\AA}$).  \textit{Middle}: The emission line at 6693.6~$\textrm{\AA}$ identified as [OIII]~5007~$\textrm{\AA}$ (EW~=~0.17~$\textrm{\AA}$),  \textit{Bottom}: The emission line at 8800.5~$\textrm{\AA}$ identified as [NII]~6583~$\textrm{\AA}$ (EW~=~0.05~$\textrm{\AA}$).   The short vertical bars indicate the fainter component of the doublet.}
\label{fig:fig3}
\end{figure}

The presence of these three emission lines is also confirmed in the observations secured at R~$\sim$~1300 resolution (see Tab.~1).
In Fig.~\ref{fig:fig3}, we reproduce the close-up of the spectral regions around the detected features. For these, we measured EW of 0.12~$\pm$~0.03~$\textrm{\AA}$, 0.17~$\pm$~0.02~$\textrm{\AA}$,  and 0.05~$\pm$~0.02~$\textrm{\AA}$, for [OII], [OIII], and [NII],  respectively.

As a consistency check, we also estimate a redshift lower limit z~$>$~0.3, based on the lack of absorption features due to the host galaxy, assuming to be a typical giant elliptical of M(R)~=~-22.9 \citep[see for details][]{paiano2017tev}.

\section{Conclusions}
We obtained an unprecedented high S/N spectrum of the BL~Lac object TXS~0506+056, that it is the likely counterpart of the IceCube neutrino event.
On the basis of three faint emission lines, we found the redshift is z~$=$~0.3365$\pm$0.0010.
At this redshift the observed $g$ luminosity is $\nu L _{\nu}$$~\sim~$7$\times$10$^{45}$  erg/s. 
Assuming that the source is hosted by a typical  massive elliptical galaxy \citep[e.g.][]{falomo2014}, the magnitude is  r(host)$\sim$ 18.9. This translates into a total nucleus to host ratio~$>$~10 that  is significantly larger than the average value found for BL Lac objects \citep[see e.g.][]{scarpa2000} and it is indicative of an highly beamed source.
The line luminosities are 2.0$\times$10$^{41}$ erg/s for the [OII] and [OIII] and 5.0$\times$10$^{40}$ erg/s for the [NII]. The [OII] luminosity is typical of what is found in QSOs \citep[see][]{kalfountzou2012}. The line ratios Log([OII]/[OIII]) and Log([NII]/[OII]) are -0.15 and -0.38, respectively. These values are consistent with what typically found for narrow line region emission lines \citep{richardson2014}.
Neither the H$_{\beta}$ , that would fall at $\lambda \sim$~6500~$\textrm{\AA}$ where a weak telluric absorption is present, nor H$_{\alpha}$ are detected in our spectra.
The optical spectrum of TXS~0506+056 therefore resembles that of a Seyfert 2 galaxy from the point of view of the emission lines. 
Based on the minimum EW in the regions of these two lines, we estimate that [OIII]/H$_{\beta}$~$\gtrsim$~3  while [NII]/H$_{\alpha}$~$\gtrsim$~2. These ratios lead to interpret that these emission lines are originated in to the narrow line region of the AGN.

\newpage

\section*{Acknowledgement}
We thank the referee for his/her prompt and constructive criticism and comments.

\facilities{GTC-OSIRIS, \citep{cepa2003}}
\software{IRAF \citep{tody1986, tody1993}}

%





\bibliographystyle{aasjournal}




\end{document}